\documentclass[a4,12pt,epsf]{article}
\usepackage{amsmath,graphicx}
\begin{document}
\begin{center}{\Large \bf
The light pseudoscalar Higgs boson in NMSSM
}
\end{center}

\begin{center}
Abdesslam Arhrib$^a$, Kingman Cheung$^b$, Tie-Jiun Hou$^b$, Kok-Wee Song$^b$
\vspace{6pt}\\

$^a$
{\it 
Facult\'e des Sciences et Techniques B.P 416 Tangier, Morocco
}
\\
$^b$
{\it  
Department of Physics and NCTS, National Tsing Hua University, Hsinchu, Taiwan
}
\end{center}
\begin{abstract}
We study the associated production of a very light pseudoscalar Higgs
boson with a pair of charginos.  The novel signature involves a pair
of charged leptons from chargino decays and a pair of photons from the
pseudoscalar Higgs boson decay, plus large missing energy at the LHC
and ILC.  The signal may help us to distinguish the NMSSM from MSSM,
provided that the experiment can resolve the two photons from the
decay of the pseudoscalar Higgs boson.
\end{abstract}


The little hierarchy problem has motivated a number of solutions to 
relieve the problem.
One of these is to add additional singlet fields to the
minimal supersymmetric standard model (MSSM).  The minimal version of the
latter is realized by adding a singlet Higgs field
to the MSSM, and becomes the next-to-minimal supersymmetric standard model
(NMSSM).  It has been shown \cite{jack} that in some corners of the parameter
space, the Higgs boson can decay into a pair of very light pseudoscalars
such that the LEP2 limit can be evaded. 
The NMSSM is in fact well motivated as it 
provides an elegant solution to the $\mu$ problem
in SUSY.  
Explicitly, the superpotential of the NMSSM is given by
\begin{equation}
W = \mathbf{h_u} \hat{Q} \, \hat{H}_u \,\hat{U}^c
- \mathbf{h_d} \hat{Q}  \, \hat{H}_d \, \hat{D}^c
- \mathbf{h_e} \hat{L}  \, \hat{H}_d \, \hat{E}^c
+\lambda \hat{S} \, \hat{H}_u \, \hat{H}_d+ \frac{1}{3}\kappa \, \hat{S}^3.
\end{equation}
The Higgs sector of the NMSSM consists of the usual two Higgs doublets 
$H_u$ and $H_d$ and an extra Higgs singlet $S$.
With the additional singlet Higgs field, there are one more CP-even and
one more CP-odd Higgs bosons, and one more neutralino  than those
in the MSSM.  
One particular feature of the NMSSM is the allowable light pseudoscalar
boson $A_1$, which is consistent with existing data.  Since this
$A_1$  mainly comes from the singlet Higgs field, it can escape all the
experimental constraints when the mixing angle with the MSSM Higgs fields
goes to zero.  It was pointed out in Ref. \cite{jack} that even in the
zero mixing limit, there is always a SM-like Higgs boson that decays into
a pair of $A_1$'s, which helps the Higgs boson to evade the LEP bound.  
The possibility to detect such light pseudoscalar Higgs bosons coming 
from the $H_1$ decay was studied using the two photon mode of the $A_1$,
but the two photons may be too collimated for realistic detection
\cite{greg}.
Here we probe another novel signature in the zero-mixing
limit \cite{km}.  The unique term $\lambda S H_u H_d$ in the superpotential
gives rise to the coupling of $\lambda S \tilde{H_u} \tilde{H_d}$,
which includes the neutral and charged Higgsinos.  We study the
associated production of a light pseudoscalar Higgs boson with a
chargino pair in the zero-mixing limit at the LHC and ILC. 
Provided that the pseudoscalar is very
light and the mixing angle is less than $10^{-3}$, the dominant decay
mode of $A_1$ is a pair of photons.  Thus, the novel signature for the
production is a pair of charged leptons and a pair of photons plus
large missing energy.  

The extra singlet field is allowed to couple
only to the Higgs doublets of the model, the supersymmetrization of which
is that the singlet field only couples to the Higgsino doublets.
Consequently, the couplings of the singlet $S$ to gauge bosons and fermions
will only be manifest via their mixing with the doublet Higgs fields. 
After the Higgs fields take on the VEV's and rotating away the Goldstone
modes, we are left with a pair charged Higgs bosons, 3 real scalar fields,
and 2 pseudoscalar fields.  The mass matrix for
 the two pseudoscalar Higgs bosons $P_1$ and $P_2$ is 
\begin{equation}
V_{\rm pseudo} = \frac{1}{2} \;( P_1 \;\; P_2)  {\cal M}^2_P \; 
           \left( \begin{array}{c}
                             P_1 \\
                             P_2 \end{array} \right )
\end{equation}
with
\begin{eqnarray}
{\cal M}^2_{P\, 11} &=& M^2_A \;,\;\;
{\cal M}^2_{P \, 12} = {\cal M}^2_{P \, 21} = \frac{1}{2} \cot\beta_s\,
\left(M^2_A \sin 2\beta - 3 \lambda \kappa v_s^2 \right ) \;, \nonumber \\
{\cal M}^2_{P \, 22} &=& \frac{1}{4} \sin 2\beta \cot^2 \beta_s
\,\left( M^2_A \sin 2\beta + 3 \lambda \kappa v_s^2 \right )
 - \frac{3}{\sqrt{2}} \kappa A_\kappa v_s \;,  \nonumber \\
M_A^2 &=& \frac{\lambda v_s}{\sin 2\beta}\left(
             \sqrt{2} A_\lambda + \kappa v_s  \right ) \label{mA}\;,
\end{eqnarray}
and $\tan \beta = v_u/v_d$ and $\tan \beta_s = v_s/v$ and $v^2=v_u^2+v_d^2$.
The pseudoscalar fields are further rotated to the diagonal basis
($A_1$, $A_2$) through a mixing angle:
\begin{eqnarray}
\left( \begin{array}{c} A_2 \\ A_1 \end{array} \right)=
\left( \begin{array}{cc}
     \cos \theta_A & \sin \theta_A \\
     -\sin \theta_A & \cos \theta_A  \end{array} \right)
\left( \begin{array}{c} P_1 \\ P_2 \end{array} \right)
\end{eqnarray}
At tree-level the mixing angle is given by
\begin{equation}
\tan \theta_A = \frac{ {\cal M}^2_{P\,12} }{ {\cal M}^2_{P\,11} - m^2_{A_1}}
 = \frac{1}{2} \cot\beta_s \,
  \frac{M^2_A \sin 2 \beta - 3 \lambda \kappa v_s^2}
  { M^2_A - m^2_{A_1} } \;.
\label{tanA}
\end{equation}
In the approximation of large $\tan\beta$ and large $M_A$,
which is normally valid in the usual MSSM, 
the tree-level CP-odd masses can be written as \cite{miller}
\begin{equation}
m_{A_2}^2 \approx M_A^2 \, (1+\frac{1}{4} \cot^2 \beta_s \sin^2 2\beta),
\qquad
m_{A_1}^2 \approx -\frac{3}{\sqrt{2}} \kappa v_s A_{\kappa}\;.
\label{small-ma}
\end{equation}
We are interested in the case that $A_1$ is very light.  From 
Eq. (\ref{small-ma}) it can be seen that $m_{A_1}$ can be very small
if either $\kappa$ or $A_\kappa$ is very small.
We can achieve a small mixing angle by the cancellation between the
two terms in the numerator of Eq.(\ref{tanA}), by setting
\begin{equation}
  M_A^2 \sin 2 \beta - 3 \lambda \kappa v_s^2 = \sqrt{2} \lambda v_s 
\left( A_\lambda - \sqrt{2} \kappa v_s \right ) \approx 0
\qquad 
 \Rightarrow A_\lambda \approx \sqrt{2} \kappa v_s \;.
\label{cond}
\end{equation}

\begin{figure}[t!]
\centering
\includegraphics[width=3in,clip]{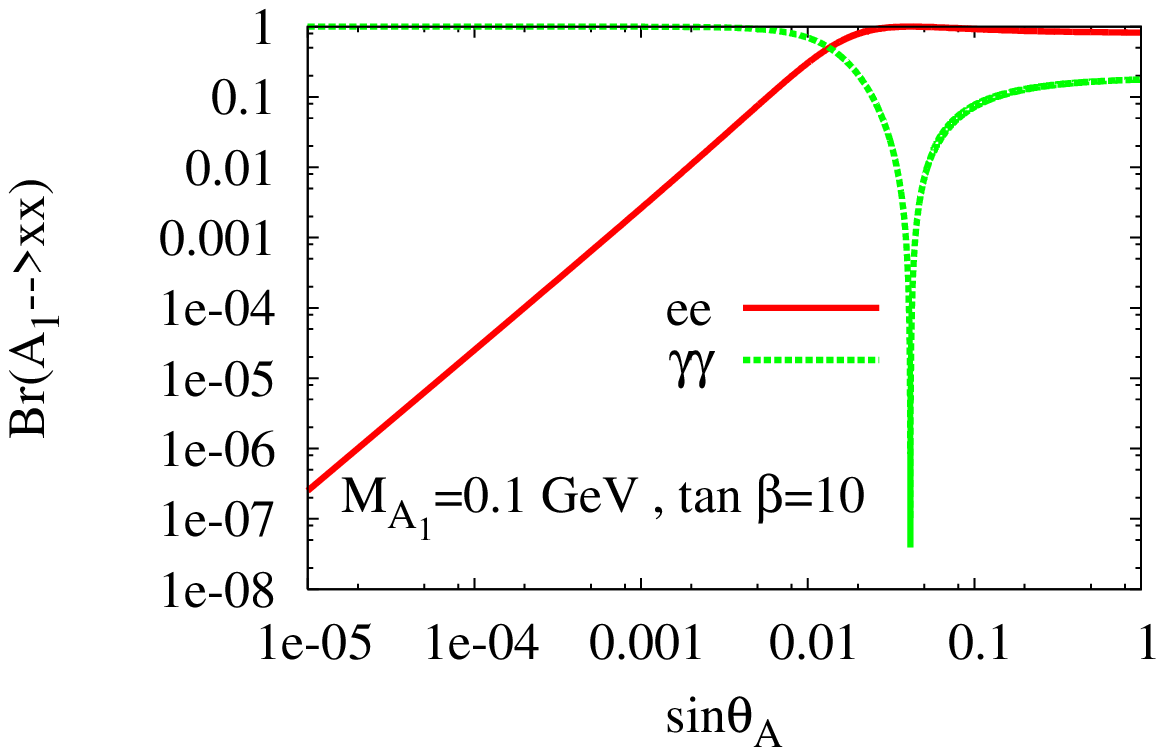}
\includegraphics[width=3in,clip]{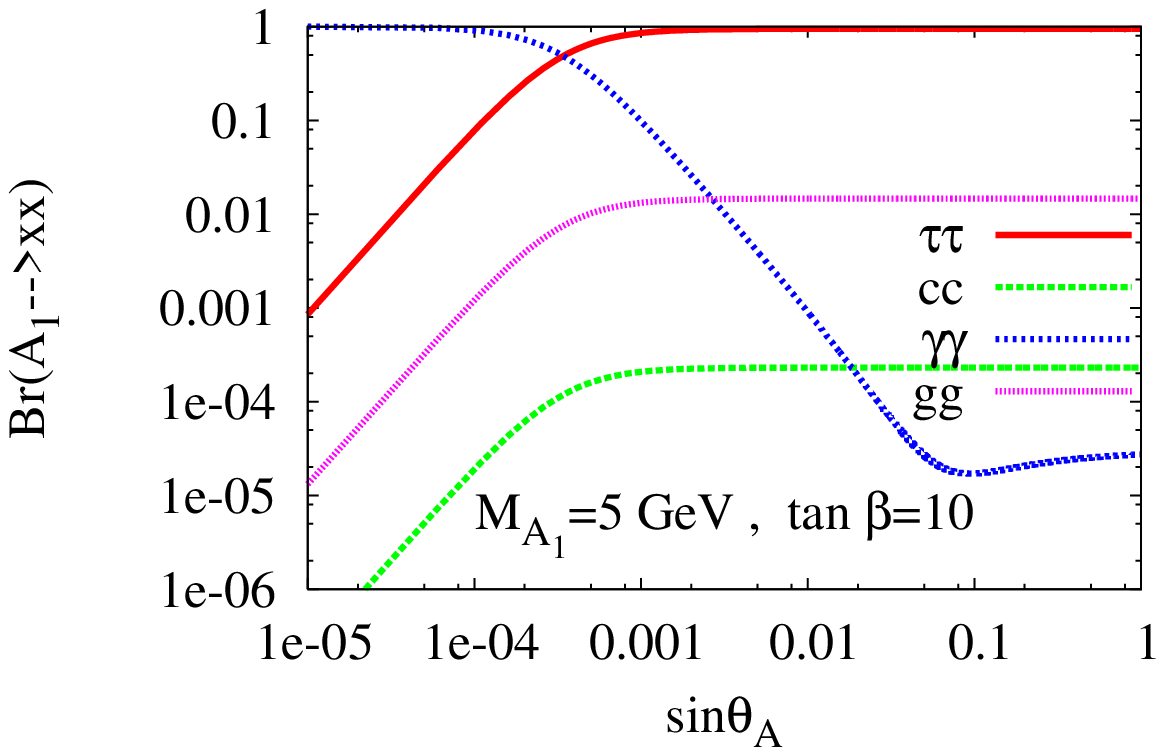}
\caption{\label{fig1} \small
Decay branching ratios for the light pseudoscalar Higgs boson
versus the mixing angle $\sin\theta_A$ for $\lambda=1$, $\mu=150$, $M_2=500$
GeV. (a) $m_{A_1} = 0.1$ GeV and (b) $m_{A_1}=5$ GeV.}
\end{figure}

In the limit of zero mixing, the $A_1$ only couples to a pair of charginos
and neutralinos.  Therefore, the dominant decay mode is
$\gamma\gamma$ via a chargino loop if $m_{A_1}$ is very light.
When we turn on the small mixing angle,
other modes, such as $q\bar q$, $\ell^+ \ell^-$, and $gg$, appear, which
will eventually dominate when the mixing angle is larger than $O(10^{-3})$.
We show a typical decay branching ratio versus the mixing angle for 
$m_{A_1} = 0.1, 5$ GeV in Fig. \ref{fig1}.
As long as $\sin\theta_A < 10^{-3}$
the $\gamma\gamma$ dominates the decay of $A_1$.  

Note that the cross section is insensitive to $\sin\theta_A$ as long as
it is less than $10^{-2}$. Also, in this near-zero mixing region, the
cross section scales as $\lambda^2$.  
The signature is very spectacular with a pair of charged leptons and 
a pair of photons with a large missing energy.  
One can also increase the detection rates by including the hadronic decay 
of the charginos.  Therefore, in the final state we can have 
(i) two charged leptons $+$ two photons $+$ $\not\!{E}_T$, 
(ii) one charged leptons $+$ two jets $+$ two photons $+$ $\not\!{E}_T$, or
(iii) 4 jets $+$ two photons $+$ $\not\!{E}_T$.

\begin{figure}[t!]
\centering
\includegraphics[width=3.5in]{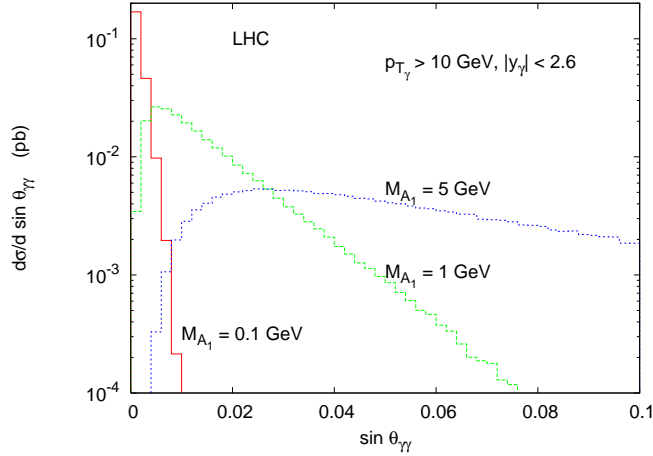}
\caption{\small \label{angle}
The differential cross section versus the sine of the opening angle
between the two photons for $\lambda = 1$ and $\sin\theta_A = 10^{-4}$ 
at the LHC.  Requirements of $p_{T\gamma} > 10$ GeV and $|y_\gamma| < 2.6$
are imposed.
}
\end{figure}

The critical issue here is whether the LHC experiment
 can resolve the two photons  in the decay of the 
pseudoscalar Higgs boson.  We perform a monte carlo study for the 
production of $\widetilde{\chi}^+_1 \, \widetilde{\chi}^-_1\, A_1$ 
followed by the decay of 
$A_1 \to \gamma \gamma$.  We impose on the photons:
\begin{equation}
p_{T_\gamma} > 10 \; {\rm GeV}\,, \qquad |y_\gamma| < 2.6 \,,
\end{equation}
which are in accord with the ECAL of the CMS detector \cite{cms}.  The 
resolution of the ``preshower'' detector
quoted in the report is as good as $6.9$ mrad.  We shall use
10 mrad as our minimum separation of the two photons.
We show the distribution of the sine of the opening
 angle between the two photons
for $M_{A_1} = 0.1,\; 1,\; 5$ GeV in Fig. \ref{angle}.  It is easy to
understand that for  $A_1$ as light as $0.1$ GeV all the cross sections are
within the opening angle $\theta_{\gamma\gamma} < 0.01$ rad.  When $M_{A_1}$ 
increases to 1 GeV, more than half of the cross sections are beyond 0.01 rad.
For $M_{A_1}$ as large as 5 GeV almost all cross sections are beyond 
$\theta_{\gamma\gamma} > 0.01$ rad.  We show the resultant cross sections
for $M_{A_1} = 0.1 - 5 $ GeV 
with $p_{T_\gamma} > 10 \; {\rm GeV}$, $|y_\gamma| < 2.6$, and 
$\theta_{\gamma\gamma}> 0.01$ rad in Table \ref{table}.  
For a $O(100)$ fb$^{-1}$ luminosity, Table I shows that it is only 
possible to detect $m_{A_1} > 1 $ GeV.

\begin{table}[t!]
\caption{\label{table} \small
Cross sections in fb for associated production of 
$\widetilde{\chi}^+_1 \; \widetilde{\chi}^-_1 \; A_1$ followed
by $A_1 \to \gamma\gamma$.  The cuts applied to the two photons are:
$p_{T_\gamma} > 10$ GeV, $|y_\gamma| < 2.6$, and $\theta_{\gamma\gamma} > 
10$ mrad.  }
\medskip
\centering
\begin{tabular}{|c|c|}
\hline
\hline
  $M_{A_1}$  ( GeV)  & Cross Section  (fb) \\
\hline
$0.1$                &  $0.0$ \\
$0.3$                &  $0.0405$  \\
$0.5$                &  $0.12$  \\
$1$                &  $0.26$  \\
$3$                &  $0.42$  \\
$5$                &  $0.44$  \\
\hline
\end{tabular}
\end{table}

There are other constraints on a light pseudoscalar from $g-2$, rare $K$ and
$B$ meson decays, $B-\overline{B}$ mixing, $B_s \to \mu^+ \mu^-$, and 
$\Upsilon \to A_1 \gamma$ \cite{hiller,hooper}.  
However, it is obvious that in these
processes the light pseudoscalar interacts via the mixing with
the MSSM pseudoscalar.  Thus, in the limit of zero-mixing the constraints
on the light $A_1$ can be easily evaded.

\vspace*{12pt}
\noindent
{\bf Acknowledgement}
\vspace*{6pt}

\noindent
The research was supported by the NSC of Taiwan under
Grant 94-2112-M-007-010-.


\end{document}